\documentstyle[aps,epsf]{revtex}
\def\tr{\,{\rm tr}\,}
\def\IR{\relax{\rm I\kern-.18em R}}
\def\eqnn#1{(\ref{#1})}

\def\cmp{{ \sl Comm. Math. Phys. }}

\def\npb{{ \sl Nucl. Phys. }}
\def\prc{{ \sl Phys. Rep. }}
\def\prd{{ \sl Phys. Rev. }}
\def\prl{{ \sl Phys. Rev. Lett. }}
\def\plb{{ \sl Phys. Lett. }}

\def\dslash{\hbox{$\partial$\kern-0.5em\raise0.3ex\hbox{/}}}
\def\pslash{\hbox{{\it p}\kern-0.48em\raise-0.3ex\hbox{/}}}
\def\half{{1\over2}}

\def\figno#1{{Fig.~\ref{fig:#1}}}
\def\intx{\int_0^1\!\!\!dx\,}

\def\vp#1{\varphi^{(#1)}}

\def\hvp{\hat\varphi}
\def\gam#1{b_{#1}}

\def\pint{P\!\!\int}
\def\nf{{N_F}}
\def\quarter{{1\over4}}
\def\bb#1{B(#1,#1)}
\def\susc{{\cal R}}
\def\vp#1{\varphi^{(#1)}}

\def\hvp{\hat\varphi}
\def\sumlk{\sum_{l=1}^K}
\def\pim#1{\left(P^{-1}M\right)_{{#1}}}
\def\beb#1{B(#1,#1)}
\draft
\begin{document}
\title{
  Mass inequalities in two dimensional gauged four fermi models
}
\author{Kenichiro Aoki\footnote{E--mail: {\tt ken@phys-h.keio.ac.jp}}
and Kenji Ito\footnote{E--mail: {\tt kito@th.phys.titech.ac.jp}}
  }
\address{$^*$Hiyoshi Dept. of Physics, Keio University, {\it
    4--1--1} Hiyoshi, Kouhoku--ku, Yokohama 223--8521, Japan\\
  $^\dagger$Dept. of Physics, Tokyo Institute of Technology,
  2--12--1 Oh-okayama, Meguro-ku, Tokyo {\it 152--0033}, Japan}
\date{\today }
\maketitle

\begin{abstract}
  We quantitatively analyze the meson mass inequality relations
  of two dimensional gauged four fermi models in the large $N$
  limit. The class of models we study includes the {}'t~Hooft
  model, the chiral and non--chiral Gross--Neveu models as
  special points in the space of field theories.  Cases where
  the chiral symmetry is spontaneously or explicitly broken are
  both studied.  We study the meson mass inequality
  quantitatively and define a susceptibility which allows us to
  systematically analyze the inequality.  In the generalized
  Gross--Neveu model limit, we derive an analytic expression for
  this susceptibility. Even though no analytic proof of the
  validity of the classic mass inequality exists for the generic
  case, the mass inequality is found to be positive throughout
  most of the parameter space.  We point out that the inequality
  might be negative in certain cases.
\end{abstract}
\section{Introduction}
\label{sec:intro}
Determining the properties of composite particles in
relativistic field theories, such as mesons or baryons in QCD,
is an essentially non--perturbative problem and is quite
difficult to address from first principles.  While impressive
progress has been made in numerical approaches to the
problem\cite{lattice}, it is highly desirable to also have a
more analytic understanding of the properties of the composite
particles.  An important relation that can be applied to composite
particles in vector like theories, like massive QCD, are the
mass inequality relations \cite{ineq}.  While these elegant
inequalities are quite useful and have therefore been well
studied (for a recent review, see, \cite{ineq-revs}), there
seems to be little understanding regarding their quantitative
behavior.  Furthermore, while the proof of inequalities does not
apply to chiral, explicitly left--right asymmetric theories or
theories with Yukawa couplings, no examples of theories wherein
the inequality has been shown to be negative exists amongst
relativistic quantum field theories.
The mass inequalities provide non-trivial insight into the
dynamics of the interacting model.  An understanding of the
behavior of the inequalities will undoubtedly further our
understanding of the spectrum of bound states in relativistic
quantum field theories.

Whether the inequalities might or might not be positive in
chiral models is of import, in particular to supersymmetric
theories.  In supersymmetric theories, one often `solves' the
gauge hierarchy problem by invoking the chiral symmetry of the
fermions, which is related in turn to the mass of the scalar
bosons by supersymmetry.  While some non--perturbative aspects
of supersymmetric gauge theories are recently being
clarified\cite{susy,nishino}, relations analogous to the mass
inequality relations seems not to be known.  Such relations, if
they exist, should shed light on the properties of the spectrum
on supersymmetric theories and on the possibility of spontaneous
breaking of supersymmetry.
In attempting to extend the mass inequalities
to supersymmetric theories, we believe a deeper understanding of
its properties to be crucial. Of course, non-supersymmetric
theories are of interest on its own, one of the reasons being
that the low energy world is not supersymmetric.

While, needless to say, computing the mass inequalities in full
QCD would be of import, not to mention very interesting, this is
a daunting task.  One approach is to compute the mass
inequalities and develop an understanding of their behavior in
analytically solvable relativistic quantum field theory models,
such as the classic {}'t~Hooft model\cite{thooft} and the
Gross--Neveu models\cite{gn}. This is the approach we shall
adopt in this work. These models are tractable yet non--trivial
and have proven to be quite instructive by providing physics
insight into the non-perturbative aspects of field
theories\cite{revs}.  Dynamics of gauge theories is certainly of
import especially since it is an integral part of the Standard
Model.  Also, theories of four fermi interactions have been
serving an important role in particle physics and other fields
of physics\cite{t-gn,topc}.

In this work, we shall analyze the properties of mass
inequalities of gauged four fermi models in $(1+1)$--dimensions,
using the large $N$ limit.  In these models, the properties of
the ``meson'' states can be reduced analytically to the problem
of solving mathematical equations\cite{burkardt,ai2}.  This will
allow us to study the problem analytically, even if the final
equations need to be solved numerically.  The parameters in this
family of models which we can arbitrary control are the scalar
and the pseudo-scalar four fermi couplings, the gauge coupling
and the ``quark'' masses.  The class of models we study contains
the {}'t~Hooft model\cite{thooft} and the chiral and non--chiral
Gross--Neveu models\cite{gn} for particular choices of the
parameters.  It includes cases where the chiral symmetry is
spontaneously or explicitly broken.  This is of particular
interest, since the Gross--Neveu model is known to be equivalent
to a Yukawa model where the scalar field develops an expectation
value\cite{gn}.  This is precisely the kind of situation where
one might suspect that the mass inequality may be
violated\cite{vw,cvetic}.

In order to study the mass inequalities quantitatively and
systematically, we need a measure for the size of the
inequality.  We shall find such a measure in a parameter which
we shall call the `meson mass susceptibility' to be explained
below.  This parameter will allow us to compare the inequalities
of various field theories in a natural manner.

The work is organized as follows:
In section~\ref{sec:susc}, we shall define and explain the meson
mass susceptibility.  In section~\ref{sec:qm}, we discuss mass
inequalities in quantum mechanics. In section~\ref{sec:gn}, we
first treat the simpler case of generalized Gross--Neveu models
and analyze the mass inequalities there.  In particular, an
analytic form of the mass susceptibility will be explicitly
presented. In section~\ref{sec:tgn}, we work with the general
gauged four fermi models. We first summarize the class of models
we study and explain how the meson spectrum can be computed.
Particular care has been paid to presenting the methods we use
explicitly for further possible use.  We then compute and study
the properties of the mass differences.  We end with discussions
of the results in section~\ref{sec:discussions}.  A short
appendix on some of the technical aspects of the computation is
included.
\section{The meson mass susceptibility}
\label{sec:susc}
In studying the mass inequalities quantitatively, we need a
quantitative measure of how ``large'' the inequality is, in
order to compare within the field theory space.  This encodes
information, intuitively speaking, on how strong the attractive
interactions in the theory are.  The mass inequality we consider
is
\begin{equation}
  \label{ineq}
  \delta\mu_{ab}\equiv 
  \mu_{ab}-{\left(\mu_{aa}+\mu_{bb}\right)\over2}
\end{equation}
which is known to be positive for vector-like gauge
theories\cite{ineq}.  Here, denoting the constituents as $q$'s,
$\mu_{ab}$ is the mass of the lightest meson that overlaps with
the $\overline{q_a}q_b$ state, and so on.

This quantity is dimensionful and depends not only on the model,
but also on the difference of the masses of the constituents.
The quantity may be made dimensionless trivially by taking the
ratio of the inequality with a meson mass.  However, this
inequality may become large just because the constituent mass
difference is large, so is not the most appropriate parameter
for investigating the intrinsic dynamics of the theory.  In
fact, if we consider the field theory space to be parameterized
by the couplings of the model which include the masses, the mass
inequality is {\it not} a local quantity in the parameter space.
It is more natural to define a local parameter in the field
theory space.  Let us define the mass squared of the
constituents to be
\begin{equation}
  \label{msq}
  M_a^2=M^2(1+\Delta),\qquad
  M_b^2=M^2(1-\Delta)
\end{equation}
The meson mass difference $\delta\mu_{ab}$ is even under the
interchange of $M_a$ and $M_b$ and is therefore an even function
of $\Delta$.

We shall characterize the inequality by a parameter we refer to
as the ``meson mass susceptibility''.  This quantity is defined
by, 
\begin{equation}
  \label{susc}
  \susc\equiv    \lim_{\Delta\rightarrow 0}
  {\delta\mu_{ab}\over\Delta^2\mu_{ab}}
  =
    \lim_{\Delta\rightarrow 0}
    {2\mu_{ab}-\left(\mu_{aa}+\mu_{bb}\right) \over
    \Delta^2 2\mu_{ab}}\qquad,
\end{equation}
which is a function of the couplings and the mass of the
constituent quark.  The susceptibility we defined above is also
useful from a practical point of view: Since the mass inequality
is an expansion in the mass difference squared, the
susceptibility together with the meson mass for equal mass
quarks, distill the meson mass information when the quark mass
difference is not too large.

We should perhaps here discuss the relation of this
susceptibility to the global properties of the mass inequality,
namely when the mass differences are arbitrary.  A natural
question is whether the positivity of the susceptibility in a
parameter region guarantees the positivity of the inequality
when the mass differences are large. In quantum mechanics, the
situation is quite simple; if the susceptibility is positive
everywhere, the mass inequality is positive for arbitrary mass
differences. This can be derived from the convexity of the meson
mass with respect to the reduced mass of the two quarks.  In
quantum field theory, however, no such argument exists in
general, since the meson mass needs not and will not depend only
on the reduced mass of the two quarks.  To make an analogous
argument in quantum field theory, we need further information
regarding the relation between the meson mass and the quark
masses.  While it seems quite natural to assume that relations
exist such that the positivity of the local susceptibility
guarantees the positivity of the mass inequality globally, we do
not know if this is in fact true.  In practice, we have found no
counterexamples to this statement.

\section{Mass inequalities in quantum mechanics}
\label{sec:qm}
In this section, we briefly discuss mass inequalities in quantum
mechanics (see also \cite{ineq-revs}).  While the discussion is
not necessary for computing mass inequalities in relativistic
field theories, we feel that it is nonetheless quite instructive
and provides a broader perspective on mass inequalities in
quantum theories.  Also, the mass inequalities in relativistic
field theories should reduce to that of quantum mechanics in the
non--relativistic limit.  As such, some of the results here will
be later compared to those from the full quantum field theory
below.  It should be noted, however, that phenomena such as
symmetry breaking which plays a large role in
$(1+1)$--dimensional gauge theories studied in this work
are essentially
quantum field theoretical so that quantum mechanical behavior is
not sufficient for understanding the full relativistic behavior,
even qualitatively.

In quantum mechanics, the problem of two body bound states under 
a local potential reduces to a model with the hamiltonian
\begin{equation}
  \label{hamiltonian}
  H = {p^2\over 2M_{12}}+V(x)
\end{equation}
where $M_{12}$ denotes the reduced mass,
$1/M_{12}=1/M_1+1/M_2$.  
We will analyze one dimensional models, but similar analysis can 
be applied to higher dimensional models.
\subsubsection{Infinitely deep square well potential}
The potential of the model is 
\begin{equation}
  V(x)=\cases{0&$(0\le x\le L)$\cr\infty& $(x<0, x>L)$\cr}. 
\end{equation}
The spectrum of the bound states is known to be
$E_{12,n}={\hbar^2\pi^2\over2M_{12} L^2}n^2,\ \ \ 
(n=1,2,\ldots)$.  This is somewhat trivial but an interesting
case.  The susceptibility $\susc=0$ and this we can understand
as the signature of the model being free within the well.
\subsubsection{Delta function potential}
The delta function potential 
\begin{equation}
  \label{delta}
  V(x)=-V_0\delta(x)\qquad (V_0>0)
\end{equation}
has a bound state with the binding energy 
$-{M_{12} V_0^2/(2\hbar^2)}$.
\begin{equation}
  \delta\mu_{ab}=E_{ab}-{E_{aa}+E_{bb}\over2}={V_0^2\over8\hbar^2}
  {\left(M_a-M_b\right)^2\over\left(M_a+M_b\right)}
\end{equation}   
This leads to the susceptibility
\begin{equation}
  \label{delta-susc}
  \susc={V_0^2\over32\left(1-V_0^2/8\right)}>0
\end{equation}
The susceptibility increases with larger $V_0$, as expected.
In the non-relativistic limit, $V_0\ll1$.
\subsubsection{Monomial potentials}
Let us also discuss potentials whose behavior is governed by a
monomial 
\begin{equation}
  \label{monomial}
    V(x)=A\,x^\gamma, \qquad A\gamma>0
\end{equation}
$\gamma$ needs not be an integer but $\gamma>-2$ needs to be
satisfied for sensible physics behavior.  $A\gamma>0$ needs to
be imposed for the existence of bound states.  $\gamma=2$ and
$-1$ corresponds to the harmonic oscillator and the three
dimensional Coulomb case, respectively.

We can use the uncertainty principle to crudely estimate the
bound state energy as 
\begin{equation}
  \label{monomial-be}
  E_{12}\simeq
  \left({\gamma\over2}+1\right)A
  \left(\hbar^2\over \gamma AM_{12}\right)^{\gamma \over \gamma +2}
\end{equation}
We can obtain the susceptibility from this energy as 
\begin{equation}
  \label{monomial-susc}
  \susc\simeq{\hbar^{2\gamma\over\gamma+2}\over8(\gamma+2)c^2}
  \left(\gamma A\over2\right)^{2\over\gamma+2}
  M^{-{2(\gamma+1)\over \gamma+2}}>0
\end{equation}
in the non--relativistic limit.  While the derivation is 
not rigorous, in the harmonic oscillator and the Coulomb cases, the
susceptibilities agrees with those obtained from exact
methods.
\section{Generalized Gross--Neveu models}
\label{sec:gn}
In this section, we analyze the mass inequalities in the
generalized Gross--Neveu models, described by the lagrangian
\begin{equation}
  \label{gn-lag}
  {\cal L} =
  \sum_{f=1}^\nf{\overline{\psi}}_f(i
  {\hbox{{$\partial$}\kern-0.52em\raise0.3ex\hbox{/}}}-m_f){\psi}_{f}
  +\frac{a^2}{2}\sum_{f,f'=1}^\nf
  ({\overline{\psi_{f'}}}{\psi_f})({\overline{\psi_f}}{\psi_{f'}})
  -\frac{a_5^2}{2}\sum_{f,f'=1}^\nf
  ({\overline{\psi_{f'}}}{\gamma}_5{\psi_f})
  ({\overline{\psi_{f}}}{\gamma}_5{\psi_{f'}})
\end{equation}
In addition to the flavor indices $f,f'$ denoted explicitly in
the above formula, the fermions carry an additional internal
space index, the `color' index ($1,2,\ldots N$) which has been
suppressed in the notation. This index should not be confused
with the flavor index.  We take the large $N$ limit while
keeping $a^2N,a_5^2N$ fixed. When $m_f=0, a_5^2=0$, the model
reduces to the original Gross--Neveu model and when
$m_f=0,a^2=a_5^2$, the model reduces to the chiral Gross--Neveu
model with continuous chiral symmetry.  When $m_f=0$ and the
couplings are not equal, we are left with discrete chiral
symmetry in the model.  We need to consider multiple flavors for
the analysis of the mass inequalities.

This class of models is included in the gauged four fermi models
we deal with below and the analytic methods discussed there can
be applied here also. However, the generalized Gross--Neveu
models can be solved completely analytically using different
methods than the gauged four fermi model case, so we
shall discuss it separately.  Here, we shall need the spectrum
in the general case when two flavors have different masses,
$m_1^2\not=m_2^2$, and $a^2\not=a_5^2$, which was not solved
explicitly in \cite{ai2}.  We shall present the spectrum and
analyze the mass inequalities.

Let us consider a meson bound state of constituents with masses,
$M_1,M_2$.  These constituent masses are physical fermion masses
that include the effects of spontaneous chiral symmetry breaking
that occurs dynamically in the Gross--Neveu model.  We dispense
with the derivation here, but the Bethe--Salpeter equation for
the meson state can be solved algebraically to obtain the meson
``wave function'', $\varphi(x)$ as
\begin{eqnarray}
  \label{gn-sol}
    \varphi(x)&=& \vp0 + \vp1 (1-2x) + \hvp(x),
  \qquad  (0\leq x\leq 1)\\
  \hvp(x)&=&{\mu_{12}^2\left(\vp0 + \vp1 (1-2x) \right)
    +2\left(M_1^2-M_2^2\right)\vp1\over 
    -\mu_{12}^2+{M_1^2\over x}+{M_2^2\over 1-x}}
\end{eqnarray}
where $\vp0,\vp1$ are constants and
$\hvp(x)/\left[x(1-x)\right]$ is integrable at $x=0,1$.  The
meson wave function satisfies the following boundary conditions
\begin{equation}
  \label{tgn-bc}
  \pmatrix{ \gam+ & (1+4G_5)\gam- \cr 
    \gam-  & (1+4G)\gam+\cr}
  \pmatrix{\vp0\cr \vp1\cr} = \intx{\hvp(x)\over x(1-x)}
  \pmatrix{ G_5 & 0 \cr  0  & G\cr}
  \pmatrix{ \gam+ & \gam- \cr  \gam-  & \gam+\cr}
  \pmatrix{1 \cr 1-2x\cr}
\end{equation}
Here, we used the notation $G\equiv a^2N/(4\pi),\ G_5\equiv
a_5^2N/(4\pi)$ for the {\it renormalized} couplings and defined
$ \gam\pm\equiv\left( M_1\pm M_2\right)/2$.  When the coupling
constants are equal, $a^2=a_5^2$,  or when the masses are
equal, $M_1=M_2$, the two boundary condition equations
simply decouple, but do {\it not} in the general case.

The boundary conditions lead to a secular equation 
\begin{equation}
  \label{gn-spectrum}
 \det\pmatrix{ J_{12} 
   -\half\left({1\over G}+{1\over G_5}\right)
   -\left({1\over G_5}-{1\over G}\right)
    {M_1^2+M_2^2\over4M_1M_2}
    &
    {M_1^2-M_2^2\over\mu^2}J_{12}+\ln{M_1^2\over M_2^2}
    -\left({1\over G_5}-{1\over G}\right)
    {M_1^2-M_2^2\over4M_1M_2}
    \cr
    -{M_1^2-M_2^2\over\mu^2}J_{12}+\ln{M_1^2\over M_2^2}
    +\left({1\over G_5}-{1\over G}\right)
    {M_1^2-M_2^2\over4M_1M_2}
    &
    \left(1-2{M_1^2+M_2^2\over\mu^2}\right)J_{12}
    -\half\left({1\over G}+{1\over G_5}\right)
    +\left({1\over G_5}-{1\over G}\right)
    {M_1^2+M_2^2\over4M_1M_2}\cr}
  =0
\end{equation}
Here, we defined
\begin{equation}
  J_{12}\equiv
  \int_0^1{\mu^2\,dx\over 
    -\mu^2 x(1-x)+M_1^2(1-x)+M_2^2 x}
\end{equation}
It should be noted that since the couplings $G,G_5$ are
dimensionless, the overall mass scale $M$ can always be scaled
out of the problem and only the relative masses have a physical
meaning. The physical parameters of this quantum field theory
are the two dimensionless renormalized couplings $G,G_5$ and the 
mass ratio $M_1/M_2$.

Before we analyze the behavior of the mass inequalities, we
first need to understand the behavior of the spectrum when the
masses of the constituents are the same.  In this case, the
secular equation \eqnn{gn-spectrum} splits into two independent
equations for the pseudo--scalar and scalar bound states, $\chi$ 
and $\sigma$:
\begin{eqnarray}
  \label{gn-sigma-pi}
    \chi:&  \quad {1\over G_5} =
  \intx {(\mu_\chi/M)^2\over 1-(\mu_\chi/M)^2 x(1-x)}
  ={4\over\sqrt{4(M/\mu_\chi)^2-1}}\tan^{-1}
   \left(1\over\sqrt{4(M/\mu_\chi)^2-1}\right)
   \\
  \sigma:& \quad {1\over G} =
  \intx {(\mu_\sigma/M)^2-4\over 1-(\mu_\sigma/M)^2 x(1-x)}
\end{eqnarray}
It should be noted that only $G_5$ ($G$) appears in the equation 
for $\chi$ ($\sigma$).

$\sigma$  exists as a non--tachyonic bound state only for $
G<-1/4$. It is not clear whether the theory is unitary for
negative $G$ and we shall consider the region $G\geq0$, so we
shall not have much more to say on $\sigma$.  The original
Gross--Neveu model corresponds to $G\rightarrow -\infty$ in our
scheme and in this limit, $\mu^2_\sigma\rightarrow 4M^2$.

\begin{figure}[htbp]
  \begin{center}
    \leavevmode
    \epsfysize=6cm\epsfbox{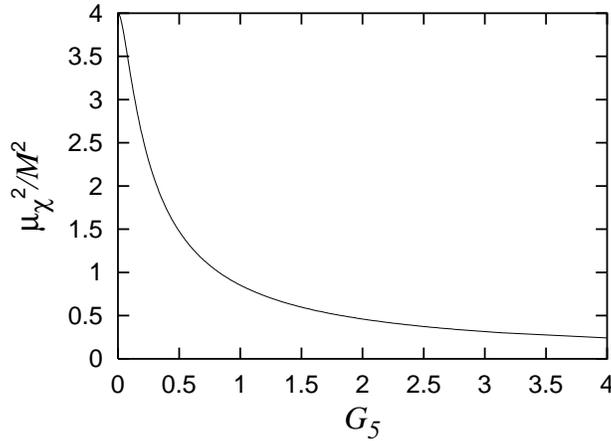}
    \caption{The behavior of the $\chi$ meson mass,
    $\mu_\chi^2/M^2$  with respect to  the coupling, $G_5$. } 
    \label{fig:gn-spectrum}
  \end{center}
\end{figure}

$\chi$ exists as a bound state for any $G_5\geq0$ and
$0\leq\mu_\chi^2\leq4M^2$. The dependence of the bound state
mass on the coupling is plotted in Fig.~\ref{fig:gn-spectrum}.
This is the only bound state in the model for $G,G_5>0$ and
corresponds to the Nambu--Goldstone like particle when the
constituent masses are zero\cite{gn,ngb}, as in the chiral
Gross--Neveu model. It is the dependence of this meson state on
the constituent masses that we shall investigate.  As a side
note, in a region we shall not investigate, there is an
intriguing possibility when $G_5\geq0$ and $G<-1/4$, in some
cases, the $\chi$ mass can be larger than the $\sigma$ mass.  We
do not know whether this can be achieved in a physically
consistent situation.  Another comment is perhaps appropriate;
in the literature, the Gross--Neveu model ($G\not=0,G_5=0$) is
often used as a prototypical simple model with a bound state.
However, the original Gross--Neveu model has no binding energy
for the meson and has barely a bound state.  It seems to us that
in fact, the simplest theory that may be considered in this
family that is useful in analyzing bound state dynamics is
$G_5\not=0, G=0$ case.  In this case, we have a bound state
whose mass depends on the coupling as in \figno{gn-spectrum}.

The meson mass susceptibility may be obtained by perturbing
the equation \eqnn{gn-spectrum} in the mass difference parameter
$\Delta$ in Eq.~\eqnn{msq}. After some computation, we derive
\begin{equation}
  \label{gn-susc}
  \susc=\left(\zeta-\quarter\right)\left\{\half
  -{1\over \left(1+{\zeta\over G_5}\right)\left[\left(\zeta-\quarter\right)
      {1\over G_5}+{1\over 4G}\right]}\right\}
\end{equation}
We defined $\zeta\equiv M^2/\mu_0^2$, where $\mu_0^2$ is the
mass of the meson in the unperturbed case, when $\Delta=0$.  The
susceptibility is independent of the overall mass scale $M$,
since it can be scaled out of the problem.  The susceptibility
may be shown analytically to be positive for any $G,G_5>0$.
Since $G, G_5$ are scalar and pseudoscalar couplings that can
take on arbitrary values, the standard proof of the mass
inequality \cite{ineq,ineq-revs} does {\it not} apply to the
models we study, except at special points.  We believe that an
analytic expression has not been derived for the mass inequality
previously in any relativistic quantum field theory.

It is interesting to check the asymptotic behavior of the
susceptibility for small and large couplings. For small $G_5$
couplings,
\begin{equation}
  \label{gn-small}
  \susc={\pi^2G_5^2\over2}\left[1-8G_5(1+4G)+{\cal O}(G_5^2)\right]
\end{equation}
This behavior is consistent with that for the $\delta$ function
problem discussed in Sect. 2.  For large $G_5$ couplings, 
\begin{equation}
  \label{gn-large}
  \susc={G_5\over2(4G+1)}-{1\over24(4G+1)^2}
  +{\cal O}\left(G_5^{-1}\right)
\end{equation}

\begin{figure}[htbp]
  \begin{center}
    \leavevmode
    \epsfysize=6cm\epsfbox{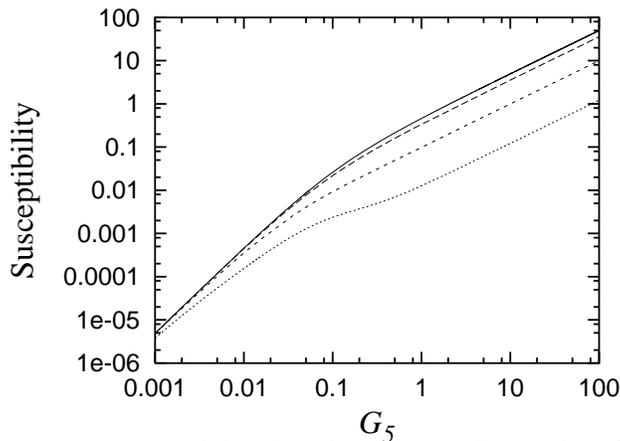}
    \caption{The behavior of the meson mass susceptibility $\cal 
      R$ with respect to the coupling $G_5$ for the generalized
      Gross--Neveu models.  The lines represent, from top to
      bottom,  $\cal R$ for $G=0,0.1,1,10$ respectively.}
    \label{fig:gn-susc}
  \end{center}
\end{figure}

The behavior of the susceptibility with respect to $G_5$ is
shown for $G=0, 0.1, 1, 10$ in \figno{gn-susc}.  The dependence
on $G$ is not strong; this is because the properties of the
bound state $\chi$ is governed mostly by the pseudo--scalar
coupling $G_5$.  The crossover from $G_5^2$ behavior to $G_5$
behavior in the susceptibility can be clearly seen in the plot.
\section{Gauged four fermi models}
\label{sec:tgn}
\subsection{The model}
\label{sec:model}
Let us now discuss the most general gauged four fermi model
described by the lagrangian 
\begin{equation}
  \label{tgn-lag}
  {\cal L} =-\frac{1}{2}\tr\left(F_{{\mu}{\nu}}F^{{\mu}{\nu}} \right)
  +\sum_{f=1}^\nf{\overline{\psi}}_f(i
  {\hbox{{\it D}\kern-0.52em\raise0.3ex\hbox{/}}}-m_f){\psi}_{f}
  +\frac{a^2}{2}\sum_{f,f'=1}^\nf
  ({\overline{\psi_{f'}}}{\psi_f})({\overline{\psi_f}}{\psi_{f'}})
  -\frac{a_5^2}{2}\sum_{f,f'=1}^\nf
  ({\overline{\psi_{f'}}}{\gamma}_5{\psi_f})
  ({\overline{\psi_{f}}}{\gamma}_5{\psi_{f'}})
\end{equation}
We have gauged the internal index in the generalized
Gross--Neveu model \eqnn{gn-lag} so that when we set the gauge
coupling to zero, we recover the generalized Gross--Neveu model
discussed in the previous section.  When we set $G=G_5=0$, we
recover the {}'t~Hooft model. We take the large--$N$ limit in a
manner similar to that of the previous section but also for the
gauge coupling; namely, we keep $g^2N,a^2N,a_5^2N$ fixed while
we take $N$ to infinity.

We again split the meson wave function as in \eqnn{gn-sol}.
Then the Bethe--Salpeter equations for the meson bound states
with the fermion constituents with masses $M_{1,2}$ may be
obtained in a simple closed form\cite{ai2}, 
\begin{eqnarray}
  \label{tgn-bs}
  \mu^2\varphi(x) &=& H\varphi(x)\nonumber\\
  &=&
  \left({\beta_1-1\over x}+{\beta_2-1\over
      1-x}\right)  \hvp(x)
  -\pint_0^1\!\!dy\,{\hvp(y)\over (y-x)^2}
  +2\vp1\left( -\beta_1+\beta_2 + \ln{1-x\over x}\right)
\end{eqnarray}
satisfying the boundary conditions \eqnn{tgn-bc}.  Since the
gauge coupling $g$ has the dimensions of the mass in
$(1+1)$--dimensions, we introduced the dimensionless mass
parameters $\beta_{1,2}\equiv\pi M_{1,2}^2/(g^2N)$.  The
renormalized couplings $G,G_5$ are defined in the same way as in
the previous section.  We should point out that all the
parameters in this equation are finite renormalized parameters,
so that the problem has been reduced to solving a somewhat
complicated integral equation. The physical parameters in this
theory are the three dimensionless renormalized parameters
$\beta,G,G_5>0$.

For later purposes, we also derive the matrix elements of the
``hamiltonian'', $H$, in the most general case, when the
couplings and the masses are arbitrary:
\begin{eqnarray}
  \label{matrix-elements}
  (\varphi',H\varphi)&=&
  \left[{\beta_1+\beta_2\over4}\left({1\over G}+{1\over G_5}\right)
    +\half\sqrt{\beta_1\beta_2}\left({1\over G_5}-{1\over G}\right)
  \right]\overline{{\vp0}'}\vp0
  +{\beta_1-\beta_2\over4}\left({1\over G}+{1\over G_5}\right)
  \left(\overline{{\vp0}'}\vp1+\overline{{\vp1}'}\vp0\right)
  \nonumber\\&&\quad
  +\left[{\beta_1+\beta_2\over4}\left({1\over G}+{1\over G_5}+8\right)+2
    +\half\sqrt{\beta_1\beta_2}\left({1\over G}-{1\over G_5}\right)
  \right]\overline{{\vp1}'}\vp1
  \nonumber\\&&\quad
  +\intx\left({\beta_1-1\over x}+{\beta_2-1\over1-x}\right)
  \overline{\hat\varphi'(x)}\hat\varphi(x)
  -P\!\int_0^1\!\!\!\int_0^1\!\!\!
  {dx\,dy\over(x-y)^2}\,\overline{\hat\varphi'(x)}\hat\varphi(y)
  \nonumber\\ &&\quad
  +\intx2\left(-\beta_1+\beta_2+\ln{1-x\over x}\right)
  \left(\overline{{\vp1}'}\hat\varphi(x)
    +\overline{\hat\varphi'(x)}\vp1\right)
\end{eqnarray}
\subsection{Methods for obtaining the spectrum}
\label{sec:tgn-spectrum}
In generalized Gross--Neveu models, the spectrum could be
obtained by just solving an ordinary equation, albeit an
transcendental one.  In contrast, for the gauged four fermi
model, we need to solve an integral equation which is
technically more involved.  Of course, this is to be expected,
since the usual {}'t~Hooft model, which is a simpler model, is
solved in terms of an integral equation.  To solve the integral
equation \eqnn{tgn-bs}, we employ two methods to be explained in
this subsection, generalizing the methods used previously in the
{}'t~Hooft model \cite{power-ref,multhopp,ai1,ai2}.  With either
of the two methods, we can solve for the spectrum and the
wavefunctions of any of the meson states for arbitrary
combinations of masses and couplings in the gauged four fermi
models.  By using the two different methods simultaneously, we
are able to obtain a better control over the error in the
results which inevitably arise when we solve the integral
equation numerically, in addition to checking the internal
consistency.  We will be succinct and summarize the results.
Even though the basic ideas are the same as those in \cite{ai2},
the results are substantially more complicated since we need to
treat the most general case which was previously not necessary.
\subsubsection{Variational method}
One method for solving the Bethe--Salpeter Eq.~\eqnn{tgn-bs},
familiar from solving the Schr\"odinger equation, is the
variational method.  We choose the basis functions
$\{\varphi_j|\ j=2,3,\ldots\}$ as
\begin{eqnarray}
  \label{basis}
  \varphi_{2k}(x)&=&c_{11}+c_{21}(1-2x)
  +{\left[x(1-x)\right]^k\over \bb k}\nonumber\\
  \varphi_{2k+1}(x)&=&c_{12}+c_{22}(1-2x)
  +{(2k+1)(1-2x)\left[x(1-x)\right]^k\over \bb k}
  \quad(k=1,2\ldots)
\end{eqnarray}
$c_{ij}$'s need to be determined to satisfy the boundary
conditions \eqnn{tgn-bc} as 
\begin{eqnarray}
  \label{cijs}
    \pmatrix{c_{11} & c_{12}\cr c_{21}&c_{22}\cr}
  &=& \pmatrix{b_+ & (1+4G_5)b_-\cr b_- & (1+4G)b_+\cr}^{-1}
   \pmatrix{G_5 & 0\cr 0 & G\cr}\pmatrix{b_+ &b_-\cr b_- &b_+\cr}
  \\ &=&
  {1\over d}\pmatrix{ 
    -{1\over4}(G-G_5)(\beta_1+\beta_2)
    +\left({G_5+G\over2}+4G_5G\right)\sqrt{\beta_1\beta_2} 
    & -{1\over4}(G-G_5)(\beta_1-\beta_2)\cr
    {1\over4}(G-G_5)(\beta_1-\beta_2) &
    {1\over4}(G-G_5)(\beta_1+\beta_2)
    +\half(G_5+G)\sqrt{\beta_1\beta_2}\cr}\nonumber
\end{eqnarray}
where
\begin{equation}
  \label{d-def}
  d\equiv (1+4G)b_+^2-(1+4G_5)b_-^2
  =\left[(G-G_5)(\beta_1+\beta_2)
    +\left(1+2(G_5+G)\right)\sqrt{\beta_1\beta_2}\right]
\end{equation}

In the variational method, the problem of obtaining the meson
states is reduced to solving an eigenvalue problem:  
\begin{equation}
  \label{tgn-variational}
  ({\mu}^2 N_{kl}-H_{kl})w_l=0,\quad 
  H_{kl}\equiv\left(\varphi_k,H\varphi_l\right),\ 
  N_{kl}\equiv\left(\varphi_k,\varphi_l\right)  \qquad
  k,l=2,3,4{\ldots}
\end{equation}
We will approximate the solution by using basis elements up to a
certain number and check the convergence by varying the
dimension of this basis space.
With some work, the matrix elements can be computed to be
\begin{eqnarray}
  \label{normalization}
  N_{2k,2l}&=&c_{11}^2+{c_{21}^2\over3}
  +{c_{11}\over2}\left({k\over2k+1}+{l\over2l+1}\right)
  +{k+l\over2(2k+2l+1)}{\beb{k+l}\over \beb k\beb l}\nonumber\\
  N_{2k+1,2l+1} &=& c_{12}^2+{c_{22}^2\over3}
  +{c_{22}\over2}\left({k\over2k+3}+{l\over2l+3}\right)
  +{(k+l)(2k+1)(2l+1)\over2(2k+2l+1)(2k+2l+3)}
  {\beb{k+l}\over \beb k \beb l}\\
  N_{2k,2l+1}&=&N_{2l+1,2k}
  = c_{11}c_{12}+{1\over3}c_{21}c_{22}+
  { k\over2(2k+1)}c_{12}+{ l\over2(2l+3)}c_{21}\nonumber
\end{eqnarray}
\begin{eqnarray}
  \label{h-variational}
    H_{2k,2l}
    &=&{1\over d}\sqrt{\beta_1\beta_2}
    \left[{1\over4}(G+G_5+8GG_5)(\beta_1+\beta_2)
      -\half(G-G_5)\sqrt{\beta_1\beta_2}\right]+    2c_{12}^2
    \nonumber\\ &&\qquad
    +(\beta_1-\beta_2)c_{12}\left(2c_{11}
      +{k\over 2k+1}+{l\over  2l+1}\right)
    +\left({\beta_1+\beta_2\over2}-1\right){\beb{k+l}\over \beb k\beb l}
    +{kl\over2(k+l)}\nonumber\\
    H_{2k+1,2l+1}&=&
    {1\over d}\sqrt{\beta_1\beta_2}
    \left[{1\over4}(G+G_5)(\beta_1+\beta_2)
      +\half(G-G_5)\sqrt{\beta_1\beta_2}\right]
    + 2c_{22}\left[c_{22}-(\beta_1-\beta_2)c_{12}\right]
    \\ &&\qquad
    +c_{22}\left({k\over k+1}+{l\over  l+1}\right)
    +\left({\beta_1+\beta_2\over2}-1\right)
    {(2k+1)(2l+1)\over 2k+2l+1}
    {\beb{k+l}\over \beb k \beb l}
    +{kl(2k+1)(2l+1)\over2(k+l)(k+l+1)}
    \nonumber\\
    H_{2k,2l+1}&=&H_{2l+1,2k}
    ={1\over 4d}\sqrt{\beta_1\beta_2}(\beta_1-\beta_2)(G+G_5)
      - 2c_{12}\left(c_{22}-(\beta_1-\beta_2)c_{12}\right)
    -c_{12}{l\over l+1}
    \nonumber\\ &&\qquad
    -c_{22}(\beta_1-\beta_2){k\over2k+1}
    +\half(\beta_1-\beta_2)
    {(2l+1) \beb{k+l}\over (2k+2l+1) \beb k \beb l}\nonumber
\end{eqnarray}
When $\beta_1=\beta_2$, the even and the odd sectors completely
decouple. 
\subsubsection{Multhopp's method}
Rather than using a variational method, we can expand the meson
wavefunction and solve the eigenvalue problem directly
\cite{multhopp}. Defining $ x\equiv (1+\cos\theta)/2$, the wave
function can be expanded in a manner consistent with the
boundary conditions as
\begin{equation}
  \label{multhopp-function} 
 \varphi(x)=
  2\pi \left(c_{11} \sum^K_{n: \rm odd} v_n
    -c_{12}\sum^K_{n: \rm even} v_n\right)
  - 2\pi \left(c_{21} \sum^K_{n: \rm odd} v_n
    -c_{22}\sum^K_{n: \rm even} v_n\right)\cos\theta
  +\sum^K_{n=1}v_n\sin n\theta
\end{equation}
where $c_{ij}$'s were defined in \eqnn{cijs}.
This reduces the Bethe--Salpeter equation \eqnn{tgn-bs} to
\begin{equation}
  \label{meson-eq-m-inf}
  \sum^K_{n=1} \left[\mu^2 \hat P_n(\theta) - \hat M_n(\theta)\right]
  v_n = 0
\end{equation}
where
\begin{eqnarray}
  \label{pmhat-def}
  \hat P_n(\theta) &\equiv&
  \sin n\theta+2\pi\cases{c_{11}-c_{21}\cos\theta& $n$: odd\cr 
    -c_{12}+c_{22}\cos\theta& $n$: even\cr}\\
  \hat M_n(\theta) &\equiv&
  2\left({\beta_1-1\over 1+\cos\theta}
    +{\beta_2-1\over 1-\cos\theta}\right)\sin n\theta
  +2\pi{n\sin n\theta\over\sin\theta}
    +    4\pi\left(\beta_1-\beta_2+
      \ln{1+\cos\theta\over1-\cos\theta}\right)
    \times\cases{-c_{21}& $n$: odd\cr c_{22} &$n$: even\cr}\nonumber 
\end{eqnarray}
The above equation \eqnn{meson-eq-m-inf} is still a functional
equation, with the dependence on the parameter $\theta$.

This can be further reduced to a generalized matrix eigenvalue
problem
\begin{equation}
  \label{multhopp-eigen}
  \left(\mu^2 P-M\right)v=0
\end{equation}
The matrices are defined as 
\begin{equation}
  \label{pm-def}
  P_{mn}\equiv\sum_{l=1}^K
  g_m(\theta_l)\hat P_n(\theta_l)
  ,\quad
  M_{mn}\equiv\sum_{l=1}^K
  g_m(\theta_l)\hat M_n(\theta_l)
  ,\qquad
  \theta_j\equiv\pi{j\over K+1}
\end{equation}
The function $g_m(\theta) $ is arbitrary, but using functions
with the property $g_m(\theta_l)=(-1)^{m+1}g_m(\theta_{K+1-l})$
simplifies the matrix elements.  With this condition, the matrix 
elements are
\begin{eqnarray}
  \label{p-elements}
 P_{mn} &= &\sumlk g_m(\theta_ l)\sin\theta_{nl}+
  2\pi\sumlk g_m(\theta_ l)
  \cases{ c_{11}& ($m,n$: odd)\cr
    c_{22}\cos\theta_ l& ($m,n$: even)\cr}
  \nonumber\\
 P_{mn}&=&-2\pi \sumlk g_m(\theta_ l)
 \cases{
  c_{12}&($m$:\ {\rm odd},$n$:\ {\rm even})\cr
  c_{21}\cos\theta_ l&($m$:\ even,$n$:\ {\rm odd})\cr}
\end{eqnarray}

\begin{eqnarray}
  \label{m-elements}
  M_{mn}&=&\sumlk g_m(\theta_ l)\left[2
    {\left(\beta_1+\beta_2-2\right)\over\sin^2\theta_ l}
    +{2\pi n\over\sin\theta_ l}\right]\sin\theta_{nl}
  +4\pi\sumlk g_m(\theta_ l)\cases{(- c_{21})(\beta_1-\beta_2) 
    & ($m,n$: odd)\cr
    c_{22}\ln{1+\cos\theta_ l\over1-\cos\theta_ l}
    & ($m,n$: even)\cr}\nonumber \\
  M_{mn}&=& -2 (\beta_1-\beta_2)\sumlk g_m(\theta_ l)
  {\cos\theta_ l\sin\theta_{nl}\over\sin^2\theta_ l}+
  4\pi\sumlk g_m(\theta_ l)
  \cases{  c_{22}(\beta_1-\beta_2)& ($m$:\ {\rm odd},$n$:\ {\rm even})\cr
    (-c_{21})\ln{1+\cos\theta_ l\over1-\cos\theta_ l}&
    ($m$:\ {\rm even},$n$:\ {\rm odd})\cr}
\end{eqnarray}
In what follows, we adopt $g_m(\theta)=2\sin\left
  [ m\theta/(K+1)\right]$ as 
was done so for the {}'t~Hooft model \cite{multhopp}.
\subsection{Mass inequalities}
\label{sec:tgn-ineq}
\begin{figure}[htbp]
  \begin{center}
    \leavevmode
    \epsfysize=6cm\epsfbox{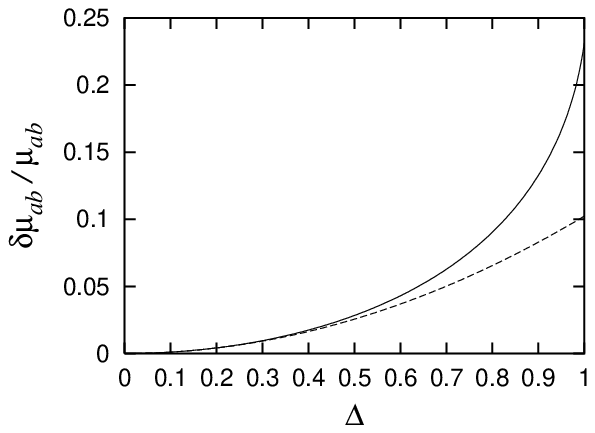}
    \caption{The normalized meson mass difference as a function
      of $\Delta$ (solid). The mass difference expected
      from the susceptibility (dashes) is also shown. } 
    \label{fig:ineqGlobal}
  \end{center}
\end{figure}

Since we have at hand the methods for obtaining the physical
properties of meson states, we are in a position to compute the
mass inequalities.  For investigating the mass inequalities, we
use the properties of the lightest meson state in each
channel. In Fig.~\ref{fig:ineqGlobal} we first plot the behavior
of the mass inequality $\delta\mu_{ab}/\mu_{ab}$ defined in
Eq.~\eqnn{ineq} for finite mass differences for a typical case
of $G=G_5=1,\ \beta=1$.  The relative quark mass difference
parameter $|\Delta|\leq1$ by definition and the mass difference
is symmetric with respect to the interchange
$\Delta\leftrightarrow -\Delta$.  At the same time, we also plot
the behavior expected from the susceptibility, ${\cal
  R}\Delta^2$.  We see that the susceptibility describes the
mass difference quite well unless the quark mass difference is
quite large, say $\Delta\gtrsim0.4$.

Let us move on to the behavior of meson mass
susceptibilities. To compute the susceptibilities, we may just
use the methods explained in the previous section and obtain the
susceptibility as the limiting case of small mass differences
going to zero.  While this is logically fine, it incurs
unnecessary numerical errors during the process. Therefore, we
can refine the method by perturbing in the mass difference {\it
  analytically} and then obtaining the mass susceptibilities
directly.  However, the standard perturbation formulas are {\it
  not} applicable to either of the two methods explained in the
previous section since we are dealing with a perturbation that
changes the boundary conditions, as we can see from
Eq.~\eqnn{tgn-bc}.  While the formulas that need to be derived
should be of use to further study, since this is technical and
somewhat involved, we have chosen to describe the methods
concretely in the appendix.  We have computed the susceptibility
using both methods and have checked that the results do agree.

The parameters of the gauged four fermi models are $\beta,G,G_5$
and the ratios of constituent masses. We expect $G$ to play a
not so dominant role in determining the properties of the
lightest meson state.  $G_5$ is the pseudo--scalar coupling that
strongly affects the lightest meson.  $\beta$ is effectively the
inverse of the strength of the gauge coupling.

\begin{figure}[htbp]
  \begin{center}
    \leavevmode
    \epsfysize=6cm\epsfbox{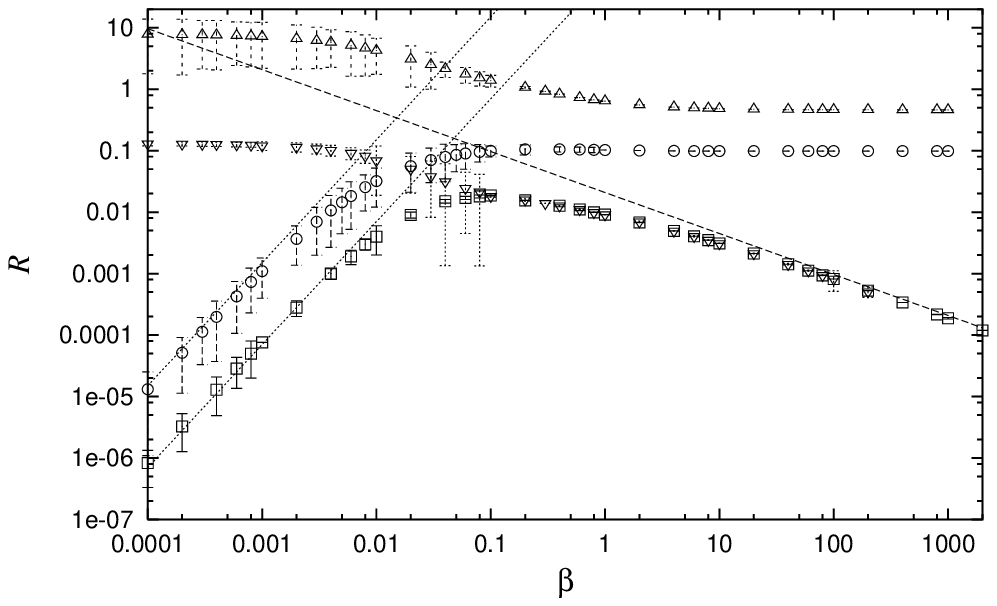}
    \caption{The behavior of $\cal R$ against $\beta$ for
      $(G,G_5)=(0,0)~(\Box),\ (1,1)~(\bigcirc),\ 
      (0,1)~(\bigtriangleup)$, and
      $(1,0)~(\bigtriangledown)$. Dashes indicate $\beta^{-2/3}
      $ behavior and dots indicate $\beta^2$ behavior.}
    \label{fig:ineq-b}
  \end{center}
\end{figure}

We first investigate the behavior of $\cal R$ with respect to
$\beta$ as in \figno{ineq-b}.  When $G_5\not=0$, for large
$\beta$, the susceptibilities approach those of the generalized
Gross--Neveu model, which is quite natural since the gauge
coupling is effectively weak. This behavior is quite visible for
$(G,G_5)=(0,1),(1,1)$ cases in \figno{ineq-b} and the approach
already occurs for moderate $\beta$ values, $\beta\gtrsim0.1$.
When $G_5=0$, as we can see from Eq.~\eqnn{gn-small}, $\susc=0$
in the generalized Gross--Neveu model.  In the gauged four fermi
model, $\susc$ behaves as $\sim \beta^{-2/3}$, when $G_5=0$ and
large $\beta$ as we can see for $(G,G_5)=(0,0),(1,0)$ cases in
\figno{ineq-b}.  This is consistent with the expectation from
the quantum mechanics calculation in Eq.~\eqnn{monomial-be} for
the linear confining potential. For $G=G_5$ and $\beta=0$, it
can be shown that ${\cal R} \rightarrow0$ as $\beta^2$. This
behavior is indeed seen in \figno{ineq-b} for
$(G,G_5)=(0,0),(1,1)$ cases.

\begin{figure}[htbp]
  \begin{center}
    \leavevmode
    \epsfysize=6cm\epsfbox{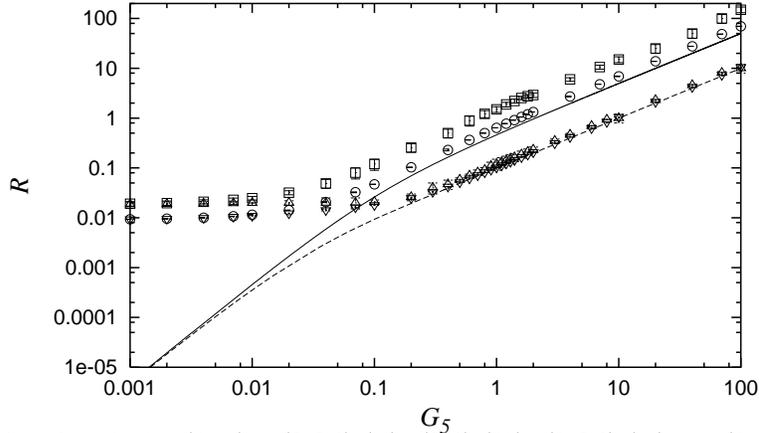}
    \caption{The behavior of $\cal R$ against $G_5$ for
      $(G,\beta)=(0,0.1)~(\Box),\ (0,1)~(\bigcirc),\ 
      (1,0.1)~(\bigtriangleup)$ and $(1,1)~(\bigtriangledown)$.
      For comparison, $\cal R$ for the generalized Gross--Neveu
      model is also displayed for $G=0$ (solid) and $G=1$
      (dashes).}
    \label{fig:ineq-g5}
  \end{center}
\end{figure}

Let us now analyze how $\cal R$ behaves with respect to $G_5$ as
in \figno{ineq-g5}.  It can be seen that for fixed $\beta$,
$\cal R$ approaches the generalized Gross--Neveu model value as
we increase $G$ or $G_5$.  Qualitatively, this can be understood
as the gauge coupling becoming relatively less important when
the other couplings are strong.  For small $G_5$, the behavior
is governed by the gauge coupling and we see in \figno{ineq-g5}
that the susceptibilities for the same $\beta$ value approach
each other.  While these behaviors can be understood from the
physics of the model as we did so here, it is quite non-trivial
derive them analytically.

\begin{figure}[htbp]
  \begin{center}
    \leavevmode
    \epsfysize=6cm\epsfbox{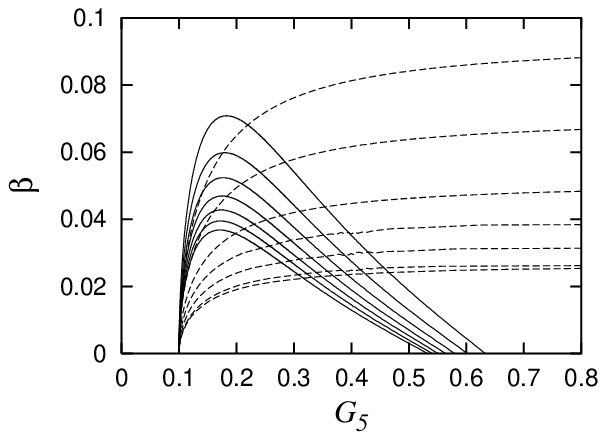}
    \caption{The zeros of the mass susceptibility parameter
      in the $G_5$--$\beta$ plane when $G=0.1$, computed using
      finite dimensional basis spaces. The solid curves, from
      top to bottom, correspond to the zeros in the variational
      method for the basis space dimensions of
      $8,10,12,14,16,18,20$.  The dashed curves, from top to
      bottom, correspond to the zeros of the susceptibility in
      the Multhopp's method for the basis space dimensions of
      $20,40,100,200,400,800,1000$. In the small regions below
      the respective curves, the susceptibility is negative. The
      negative region becomes smaller with the increase in the
      size of the basis space in both methods.}
    \label{fig:zeros}
  \end{center}
\end{figure}

We have investigated the susceptibility extensively within the
parameter space of the theory and found that it is positive,
except for a relatively small region, which we now discuss. In
most regions in the parameter space, the numerical convergence
of the susceptibility is quite rapid at least in one of the
methods and both methods yield consistent results.  In all these
regions, the susceptibility parameter satisfies ${\cal
  R}>0$. However, for small $\beta$, the convergence is rather
slow.  Particularly intriguing is the region $G\gtrsim G_5,
0\lesssim\beta\ll1$.  A simple argument shows why the behavior
in this region can be subtle: In general, the finite dimensional
numerical results are analytic with respect to the parameters
$G,G_5,\beta$. From the behavior ${\cal R}\rightarrow0$ as
$\beta\rightarrow0, G=G_5$, we know that unless $\cal R$
vanishes as $(G-G_5)^2$ or some higher even power for $\beta=0$,
there will be a region where $\cal R$ is negative.  Indeed, we
find in the numerical results that $\cal R$ is negative in the
region $G\gtrsim G_5,0\lesssim\beta\ll1$ using both the
variational and the Multhopp's method. Even the extrapolated
values, in some cases, are negative.  The meson mass squared is
always positive even in these cases and the physics of the
system seems to be quite consistent.  Naively, we would
claim that the susceptibility and hence also the mass inequality
is negative in this regime. This, to our knowledge, would {\it
  not} conflict with any general arguments regarding mass
inequalities.  However, to conclude this would be somewhat
premature, since if we study the negative region in the
parameter space, we find as in \figno{zeros} that it shrinks
when the basis space is enlarged and the negative region is
quite small compared to ${\cal O}(1)$ which is the natural scale
in the problem.  It should also be noted that even if ${\cal R}
=a_1(G-G_5)+{\cal O}((G-G_5)^2)$ for $\beta=0$ when the basis
space is finite, it is still possible that the coefficient $a_1$
approaches $0$ in the full basis space, so that the positive
susceptibility is compatible with a negative one in the
truncated basis space.  On the other hand, the regions of
negative susceptibility have a common region with respect to
both methods so it is also possible that a finite region of
negative susceptibility remains even when the basis space is
complete. We therefore conclude that while the susceptibility
may be negative in the regime $G\gtrsim G_5,0\lesssim\beta\ll1$,
further investigation is necessary to clarify this point.  An
analytic computation determining the sign of the inequality
would be ideal.  If this is not possible, a set of basis
optimized for the gauged four fermi models in this particular
parameter regime, in either variational or Multhopp's method
should settle this issue.
\section{Summary and discussions}
\label{sec:discussions}
We have systematically and  quantitatively studied the mass
inequalities in gauged 
four fermi models in $(1+1)$--dimensions. Even in the cases
where the mass inequality 
has been shown to be positive from general arguments, the size
of the inequality is unknown unless an explicit computation is
made.  We believe that the mass inequality is an interesting
dynamical quantity characterizing the spectrum of relativistic
quantum field theories. To analyze the inequalities
quantitatively, we adopted a natural susceptibility parameter to
compare the size of the inequalities throughout the field theory
space. We found that the parameter captures the essence of the
mass inequality when the constituent mass differences are not
too large.  In the family of generalized Gross--Neveu
models, we were able to derive an analytic expression for the
meson mass susceptibility.  In the more general case of gauged
four fermi models, we have developed methods for obtaining the
mass inequalities systematically and have computed them.
Since not much seems to be known about the
quantitative behavior of mass inequalities, we think that it is
significant to have a 
class of relativistic field theory models where it has been
studied explicitly.  While the results are interesting,
there remain further questions which should be answered.

An important question is whether the positivity of the mass
inequality is much more general than the cases where it has been
shown to hold \cite{ineq}.  In particular, an intriguing problem
is whether there is a relativistic quantum field theory wherein
the mass inequality is negative yet its physics behavior is
consistent. We have found that the meson mass susceptibility is
positive for most of the parameter space in the gauged four
fermi models and have explained some of the behavior
analytically.  The models we studied here include the
celebrated models of {}'t~Hooft and of Gross and Neveu for
special choices of the parameters. For the {}'t~Hooft model, the
standard arguments\cite{ineq} {\it do} apply and we may show
analytically that the inequality is positive.  However, in
general, no such arguments can be applied to gauged four fermi
models.  Furthermore, the Gross--Neveu models are known to be
equivalent to models with Yukawa couplings and also display dynamical 
chiral
symmetry breaking behavior.  These are exactly the kind of
situations in which we might doubt that the mass inequality is
positive\cite{vw,cvetic}. It is interesting that even in these
cases, the mass inequality is positive, so that in fact, the
property holds in much more general than those situations where
it has been proven.  It would be interesting to find an analytic
proof for this property if possible.  It is important to
understand why the inequality is positive for the gauged four
fermi models and clarify if this can be extended to other
theories, such as supersymmetric theories with bound states.
There still remains a small region within the field theory space
wherein the sign of the susceptibility, hence also the mass
inequality, remains uncertain and further investigation is
necessary to establish its sign.  While $(1+1)$--dimensional
theories such as the {}'t~Hooft model or the Gross-Neveu model
have physical behavior resembling those of higher dimensions, we
should mention the possibility that in higher dimensions, the
behavior of the susceptibility might be quite different.  Also,
even in $(1+1)$--dimensions, the mass inequalities might behave
qualitatively differently for other class of models.

We have seen in section~\ref{sec:qm} that the mass inequality is
satisfied in a large class of quantum mechanics
models\cite{ineq-revs}.  This leads us to suspect that the mass
inequality is valid for a large class of relativistic quantum
field theories also.  This is certainly consistent with our
findings here.  However, it should be noted that spontaneous
breaking of symmetries is essentially a field theoretical
behavior, which is also quite relevant to the theory we studied.
Therefore, we believe that it would be worthwhile to perform
further research and in particular, clarify whether the mass
inequality can become negative in relativistic quantum field
theories.  In another direction, large $N$ limit of field
theories, such as the class of models we study, are presumably
described by some kind of string theories\cite{thooft} --- an
idea, which has recently been made more concrete\cite{susy}.  It
would be interesting to find out what kind of string theories
our models correspond to and to elucidate how mass inequalities
fit into the string picture.
\appendix
\section{Perturbation theory in the mass differences for the spectrum}
\label{sec:pt}
Here, we shall briefly outline how to perform perturbation with
respect to the relative constituent mass difference, $\Delta$,
in the methods explained in section~\ref{sec:tgn-spectrum} for
obtaining the spectrum. The standard perturbation methods can
not be applied here.  One major reason is that the boundary
conditions \eqnn{tgn-bc} depend on the masses of the
constituents so that they need to be perturbed also.  There are
additional complications for both the methods used in
section~\ref{sec:tgn-spectrum}, as we shall describe below.

In both cases, we perturb in the relative mass difference
$\Delta$ and obtain an expansion for the meson mass in terms of
$\Delta$, for the cases
$(M_1^2,M_2^2)=(M_a^2,M_a^2),(M_b^2,M_b^2),(M_a^2,M_b^2)$. 
\begin{equation}
  \label{mu-exp}
  \mu^2\equiv \mu^2_0+\Delta \mu^2_1 +
  \Delta^2\mu^2_2+{\cal O}(\Delta^3)
\end{equation}
In the first two cases, the first order term exists and are of
the same size but of opposite sign and in the last case the
first order term is absent.  Therefore, the leading order term
in the mass difference $\delta\mu_{ab}$ will be of order
$\Delta^2$, as it should be.
\subsubsection{Variational method}
In the variational method, we need to consider a generalized
eigenvalue problem with the normalization matrix not being the
identity matrix.  In theory, we can just orthonormalize the
basis vectors, but in practice, this is not numerically
equivalent since the normalization matrices can become almost
singular even though we have tried to normalize the matrix
elements to be of order one.  Furthermore, since the boundary
conditions also are perturbed, the normalization matrices will
also have a non--trivial expansion in $\Delta$.

Let us expand the matrices as
\begin{equation}
  \label{v-exp}
  H\equiv H_0+\Delta H_1 + \Delta^2H_2+{\cal O}(\Delta^3),\qquad
  N\equiv N_0+\Delta N_1 + \Delta^2N_2+{\cal O}(\Delta^3)
\end{equation}
Assume that we have the complete eigen system for the 0--th
order problem:
\begin{equation}
  \label{v-0}
  H_0w_{0n}=\mu^2_{0n}N_0 w_{0n},\qquad
  \left(w_{0m},N_0w_{0n}\right)=\delta_{mn}
\end{equation}
Then, we obtain the expansion for mass squared of the meson
state labeled by $n$ 
\begin{eqnarray}
  \label{v-result}
  \mu^2_{1n}&=&\left(w_{0n},
    \left(H_1-\mu^2_{0n}N_1\right)w_{0n}\right) \nonumber\\
  \mu^2_{2n}&=&\left(w_{0n},
    \left(H_2-\mu^2_{0n}N_2\right)w_{0n}\right) 
  -\left(w_{0n},N_1w_{0n}\right)
  \left(w_{0n},\left(H_1-\mu^2_{0n}N_1\right)w_{0n}\right)\\
  &&\qquad
  +\sum_m{1\over\mu^2_{0n}-\mu^2_{0m}}
  \left|\left(w_{0n},\left(H_1-\mu^2_{0n}N_1\right)w_{0m}
    \right)\right|^2\nonumber
\end{eqnarray}
We need the expansions of the matrices $H,N$ in terms of
$\Delta$ for the three cases,
$(M_1^2,M_2^2)=(M_a^2,M_a^2),(M_b^2,M_b^2),(M_a^2,M_b^2)$, to
obtain the final results.  Since this expansion is cumbersome
but logically straightforward, it will not be explicitly
presented here to save space.
\subsubsection{Multhopp's method}
In Multhopp's method, the matrices are not Hermitean so that we
need to perform the perturbation theory with some care.
Furthermore,  due to the perturbation in the boundary
conditions, the matrix $P$ will also be perturbed. To perform
the expansion, we will reduce the equation to a mathematically
equivalent problem, 
\begin{equation}
  \label{multhopp-easy}
  \left(\mu^2-P^{-1}M\right)v=0
\end{equation}
Ideally, it is better not to invert matrices numerically, but it
is a substantially more complicated numerical task to solve a
generalized non--symmetric eigenvalue problem and also, in this
case, the matrix $P$ turns out to be quite robust against
inversion even for moderately large basis spaces with dimensions
of order $10^3$.

We expand the matrices as 
\begin{eqnarray}
  \label{multhopp-exp}
  P&=&P_0+\Delta P_1+\Delta^2 P_2,\qquad
  M=M_0+\Delta M_1+\Delta^2 M_2\nonumber\\
  \left(P^{-1}M\right)&=&\left(P^{-1}M\right)_0
  +\Delta  \left(P^{-1}M\right)_1
  +\Delta^2\left(P^{-1}M\right)_2+{\cal O}(\Delta^3)
\end{eqnarray}
where
\begin{eqnarray}
  \label{pm-exp}
    \left(P^{-1}M\right)_0 &=&P_0^{-1}M_0,\qquad
  \left(P^{-1}M\right)_1 =P_0^{-1}\left(M_1-P_1P_0^{-1}M_0\right)\nonumber\\
  \left(P^{-1}M\right)_2 &=&P_0^{-1}\left(M_2-P_2P_0^{-1}M_0
    -P_1P_0^{-1}M_1+P_1P_0^{-1}P_1P_0^{-1}M_0\right)
\end{eqnarray}

We need to first solve the $0$--th order problem for the left
and right eigenvectors, $\{u_n\}$ and $\{v_n\}$\footnote{From a
  mathematical point of view, additional complications can arise
  in general; namely the eigenvalues may be degenerate so that
  the matrix is not diagonalizable, or the eigenvalues may be
  complex. However, we need to keep in mind that we do not have
  to solve the problem for general dimensions of the basis
  space, but only for a sequence of spaces that will allow us to
  obtain the susceptibility.  In practice, these complications
  do not hinder our computations in the cases we have
  studied.}
\begin{equation}
  \label{multhopp-0}
  u_{0m}\pim0=\mu^2_{0m}u_{0m},\qquad
 \pim0v_{0n}=\mu^2_{0n}v_{0n},\qquad
 \left(u_{0m},v_{0n}\right)=\delta_{mn}  
\end{equation}
Then, we may obtain the expansion for the meson mass squared
of the meson state labeled by $n$ as 
\begin{eqnarray}
  \label{multhopp-mu}
  \mu_{1n}^2&=& \left(u_{0n},\pim1v_{0n}\right)\\
  \mu_{2n}^2&=&
  \left(u_{0n},\pim2v_{0n}\right)
  +\sum_{k\not=n}{\left(u_{0n},\pim1v_{0k}\right)
  \left(u_{0k},\pim1v_{0n}\right)\over
  \mu_{0n}^2-  \mu_{0k}^2}
\end{eqnarray}
The rest proceeds as in the variational method case.  As in the
case of the variational method, the explicit expressions for the
matrices are not shown here due to space considerations.


\begin{references}
\bibitem{lattice} For instance, see the contributions in 
  {\sl Nucl.\ Phys.\ Proc.\ Suppl.\ } {\bf B94} (2001) 
\bibitem{ineq} D. Weingarten, \prl{\bf51} (1983) 1830; 
  E.~Witten, \prl{\bf 51} (1983) 2351; S.~Nussinov, \prl{\bf 51}
  (1983) 2081
\bibitem{ineq-revs} S. Nussinov,  M.A. Lampert, 
  hep-ph/9911532 
\bibitem{susy}
N.~Seiberg and E.~Witten,
{\sl Nucl.\ Phys.}\  {\bf B426} (1994) 19,
{\sl Nucl.\ Phys.\  }{\bf B431} (1994) 484;
K.~Intriligator and N.~Seiberg,
{\sl Nucl.\ Phys.\ Proc.\ Suppl.}\  {\bf 45BC} (1996) 1;
J.~Maldacena,
{\sl Adv.\ Theor.\ Math.\ Phys.}\  {\bf 2} (1998) 231
S.~S.~Gubser, I.~R.~Klebanov and A.~M.~Polyakov,
{\sl Phys.\ Lett.}\ B {\bf 428} (1998) 105;
E.~Witten,
{\sl Adv.\ Theor.\ Math.\ Phys.}\  {\bf 2} (1998) 253;
O.~Aharony, S.~S.~Gubser, J.~Maldacena, H.~Ooguri and Y.~Oz,
{\sl Phys.\ Rept.}\  {\bf 323} (2000) 183;
\bibitem{nishino}H. Nishino, \prd{\bf D61} (2000) 025008
\bibitem{thooft}  G. {}'t Hooft, \npb{\bf B72} (1974) 461,
  \npb{\bf B75} (1974) 461 
\bibitem{gn} D.J. Gross, A. Neveu, \prd{\bf D10} (1974) 3235
\bibitem{vw}C. Vafa, E. Witten, \npb{\bf B234} (1984) 173
\bibitem{cvetic}M. Cvetic, \npb{\bf B279} (1987) 593;
  U. Maryland preprint MDDP-PP-85-023
  (1984) 
\bibitem{revs} S. Coleman, {\sl ``Aspects of symmetry''},
  Cambridge University Press (1985);\hfil\break
  B. Rosenstein, B.J. Warr, S.H. Park, \prc {\bf C205} (1991) 59
\bibitem{t-gn} Y. Nambu, G. Jona-Lasinio \prd{\bf122} (1961)
  345;\hfil\break 
  For recent reviews, see for instance, 
  J.~L.~Rosner,
  {\sl Comments Mod.\ Phys.\ } {\bf A1} (1999) 11; 
  R.S. Chivukula, {\tt hep-ph/9903500}. 
\bibitem{topc}
  Y. Nambu, in {\sl New Theories in Physics}, Z. Ajduk,
  S. Pokorski, A.~Trautman (eds), World Scientific, Singapore
  (1989); 
  and in {\sl New Trends in Strong Coupling Gauge
  Theories},  M. Bando, T. Muta, K. Yamawaki (eds), World
  Scientific, Singapore (1989);\hfil\break
  V.A. Miransky, M. Tanabashi, K. Yamawaki, {\sl
  Mod. Phys. Lett. }{\bf A4} (1989)
  1043; \plb{\bf B221} (1989) 177;\hfil\break
  W.A. Bardeen, C.T. Hill, M. Lindner, \prd{\bf D41} (1990) 1647 
\bibitem{burkardt} M. Burkardt, \prd{\bf D56} (1997) 7105
\bibitem{itakura} K. Itakura, Ph D thesis, Tokyo University
  (1996) 
\bibitem{ai1} K. Aoki, T. Ichihara, \prd{\bf D52} (1995) 6435
\bibitem{ai2}   K.Aoki,  K.Ito, \prd{\bf D60} (1999) 096004
\bibitem{ngb} S. Coleman, \cmp{\bf 31} (1973) 259
\bibitem{power-ref} W.A. ~Bardeen,
  R.B. Pearson, E. Rabinovici, \prd{\bf 21} (1980) 1037
\bibitem{multhopp} 
  A.J.~Hanson, R.D. Peccei, M.K. Prasad, \npb{\bf B121} (1977) 477;\hfil\break
  R.C.~Brower, W.L.~Spence, J.H. Weis, \prd{\bf 19D} (1979) 3024;\hfil\break
  S. Huang, J.W.~Negele, J.~Polonyi, \npb{\bf B307} (1988) 669;\hfil\break
  R.L.~Jaffe, P.F. Mende, \npb{\bf B369} (1992) 182
\end{references}
\end{document}